\RequirePackage[2020-02-02]{latexrelease}
\documentclass[%
aps, 
prl, 
reprint,
superscriptaddress,
preprintnumbers,
nofootinbib,
nobibnotes,
amsmath,amssymb,
]{revtex4-2}

\usepackage{graphicx}
\usepackage{dcolumn}
\usepackage{color}
\usepackage{lipsum}
\usepackage{bm}
\usepackage[dvipsnames]{xcolor}
\usepackage{CJKutf8}

\usepackage[colorlinks = true,
            linkcolor = blue,
            urlcolor  = blue,
            citecolor = blue,
            anchorcolor = blue]{hyperref}

\definecolor{RedWine}{rgb}{0.743,0,0}

\definecolor{NavyBlue}{rgb}{0,0,0.9}

\definecolor{DarkGreen}{rgb}{0,0.277 ,00}
\newcommand{\at}[1]{\textcolor{DarkGreen}{\bf #1}}
\definecolor{FluorescentPink}{rgb}{1.0, 0.08, 0.58}

\usepackage{url}

\newcommand{\be}{\begin{equation}}
\newcommand{\ee}{\end{equation}}
\newcommand{\bea}{\begin{eqnarray}}
\newcommand{\eea}{\end{eqnarray}}
\newcommand{\vs}{\nonumber\\} 
\def\ba#1\ea{\begin{align}#1\end{align}}

\renewcommand{\(}{\left(}
\renewcommand{\)}{\right)}

\def\mnras{Mon.\ Not.\ R.\ Astron.\ Soc.}

\def\aap{Astron.\ Astrophys.}
\def\apj{Astrophys.\ J.}
\def\aj{Astron.\ J.}
\def\apjs{Astrophys.\ J. Supp.}

\def\prd{Phys.\ Rev.\ D}

\def\physrep{Phys. Rep.}

\def\jcap{{JCAP\ }}
\def\pasj{PASJ}
\def\apss{Astrophysics and Space Science}

\newcommand{\refeq}[1]{Eq.~(\ref{eq:#1})}

\newcommand{\reffig}[1]{Fig.~\ref{fig:#1}}

\begin{document}
\preprint{KIAS-P22050, YITP-22-80}

\title{Perturbation Theory Remixed: Improved Nonlinearity Modeling beyond Standard Perturbation Theory}

\author{Zhenyuan Wang (\begin{CJK*}{UTF8}{gbsn}王震远\end{CJK*})}
\affiliation{Department of Astronomy and Astrophysics and Institute for Gravitation and the Cosmos, 
The Pennsylvania State University, University Park, PA 16802, USA}

\author{Donghui Jeong}
\affiliation{Department of Astronomy and Astrophysics and Institute for Gravitation and the Cosmos, 
The Pennsylvania State University, University Park, PA 16802, USA}
\affiliation{School of Physics, Korea Institute for Advanced Study, Seoul, South Korea}

\author{Atsushi Taruya}
\affiliation{Center for Gravitational Physics and Quantum Information, Yukawa Institute for Theoretical Physics, Kyoto University, Kyoto 606-8502, Japan}
\affiliation{Kavli Institute for the Physics and Mathematics of the Universe, Todai Institutes for Advanced Study, The University of Tokyo, Kashiwa, Chiba 277-8583, Japan}

\author{Takahiro Nishimichi}
\affiliation{Center for Gravitational Physics and Quantum Information, Yukawa Institute for Theoretical Physics, Kyoto University, Kyoto 606-8502, Japan}
\affiliation{Kavli Institute for the Physics and Mathematics of the Universe, Todai Institutes for Advanced Study, The University of Tokyo, Kashiwa, Chiba 277-8583, Japan}

\author{Ken Osato}
\affiliation{Center for Frontier Science, Chiba University, Chiba 263-8522, Japan}
\affiliation{Center for Gravitational Physics and Quantum Information, Yukawa Institute for Theoretical Physics, Kyoto University, Kyoto 606-8502, Japan}

\date{\today}

\begin{abstract}
We present a novel $n$EPT ($n$th-order Eulerian Perturbation Theory) scheme to model the nonlinear density field by the summation up to $n$th-order density fields in perturbation theory. The obtained analytical power spectrum shows excellent agreement with the results from all 20 Dark-Quest suites of $N$-body simulations spreading over a broad range of cosmologies. The agreement is much better than the conventional two-loop Standard Perturbation Theory and would reach out to $k_{\rm max}\simeq 0.4~h/{\rm Mpc}$ at $z=3$ for the best-fitting \textit{Planck} cosmology, without any free parameters. The method can accelerate the forward modeling of the non-linear cosmological density field, an indispensable probe of cosmic mysteries such as inflation, dark energy, and dark matter.
\end{abstract}


\maketitle


{\it Introduction} ---
The observations of the Universe's large-scale structure (LSS) traced by Cosmic Microwave Background (CMB) radiation \cite{WMAP9, aghanim2020planck}, the distribution of galaxies \cite{2df:2001,sdss:dr1,wigglez:2010,BOSS:2013,eBOSS:2016}, and shape distortion of galaxies \cite{KiDS:2017,DES:2018} have led to the concordance $\Lambda$CDM cosmology \cite{Planck:cosmology} with most parameters measured to a sub-percent accuracy. 

Parallel to the data collection programs has been the theoretical development based upon which we interpret the observation. Starting from the 40s \cite{Lifshitz:1945}, relativistic theory for the evolution of density perturbations in the Friedmann--Lema\^{i}tre--Robertson--Walker Universe has been developed and used to interpret the LSS data. In particular, its linearized version \cite{Peebles/Yu:1970, Sunyaev/Zeldovich:1970, Bond/Efstathiou:1987} has been so successful in explaining the power spectrum of CMB temperature anisotropies and polarizations to all scales observed by WMAP \citep{WMAP9:cosmology} and \textit{Planck} \citep{Planck:cosmology} satellites. The concordance $\Lambda$CDM cosmology model would not be possible without such an accurate linear-theory model.

The remaining big questions in cosmology are to uncover the nature of building blocks of the $\Lambda$CDM cosmology, such as inflation, dark energy, and dark matter. To address these questions, modern galaxy surveys are mapping the distribution and shape distortion of galaxies with unprecedented depth and volume \citep{hill2008hetdex, takada2014subaru, levi2013desi, lsst2012lsst,  amendola2018Euclid, maartens2015SKA, dore2014spherex}.

These observational developments call for a novel theoretical model beyond the linear theory that is only applicable on large scales where the accuracy of the usual LSS observation is limited by cosmic variance. Using a feature such as Baryon Acoustic Oscillation that is insensitive to the nonlinearities has proven successful for measuring the geometry of the Universe \citep{beutler20116df, alam2017clustering, ross2015clustering}. Upon modeling nonlinearities, however, using the full power-spectrum shape can improve the measurement accuracy by a factor of few \cite{Shoji/etal:2009, philcox2020combining}. In addition, the full-shape analysis enables the measurement of the growth rate of the LSS \citep{Kobayashi/etal:2022} and features in the galaxy clustering carved by massive neutrinos \citep{lesgourgues2006massive} and primordial physics \citep{philcox/ivanov:2022}.

A diversity of modeling methods have been developed ranging from simulation-based methods such as emulator \citep{Mira-Titan:2016,Aemulus:2019,DarkQuest:2019,BACCO:2021}, fast simulations \citep{COLA:2013,PINOCCHIO:2013,ALPT:2013,EZmock:2015}, and machine learning \citep{he2019learning,CAMELS:2021} to analytical methods such as Standard Perturbation Theory (SPT) \citep{bernardeau2002large,jeong2006perturbation}, Lagrangian Perturbation Theory (LPT) \citep{bernardeau2002large, white2014zel, chen2021redshift}, Effective-Field Theory of Large Scale Structure (EFTofLSS) \citep{baumann2012cosmological,Carrasco:2012,Hertzberg:2014}, and various Renormalized Perturbation Theory (RPT) \citep{crocce2006renormalized, Crocce_Scoccimarro2006, Crocce_Scoccimarro2008, Valageas2007, Taruya_Hiramatsu2008, Taruya_Nishimichi_Saito_Hiramatsu2009, Matsubara2008, Pietroni2008, Bernardeau_Crocce_Scoccimarro2008, Bernardeau_Crocce_Scoccimarro2012, Blas_etal2016a, Blas_etal2016b}, including RegPT \cite{taruya2012direct,osato2019perturbation}. 

Traditionally, the analytical methods focus on obtaining the expressions for the ensemble mean of the summary statistics such as the power spectrum and bispectrum, or $n$-point correlation functions. Such expressions usually involve high-dimensional integrals whose complexity increases quickly for the higher-order loop calculations. While Refs.~\cite{Schmittfull/etal:2016,mcewen2016fast,fang2017fast,Simonovic/etal:2018,Osato/etal:2021} have developed fast methods for computing nonlinear power spectrum and bispectrum by using {\sf FFTlog} algorithm \cite{Hamilton:2000} or response function expansion \cite{taruya2012direct,Nishimichi/etal:2017}, the analytical computation beyond the two-loop proves challenging \cite{Schmittfull/Vlah:2016}.

The PT-based analytical methods can also be used to model the cosmic density field at the field level \cite{Baldauf/etal:2016,TNJ2018,Schmidttfull/etal:2019,TNJ2021,schmidt2021nLPT,Schmidttfull/etal:2021,TNJ2022}. Instead of computing the ensemble mean of the summary statistics, the field-level computation provides nonlinear density fields from a given realization of the stochastic linear field. In this method, the computation of higher-order summary statistics is much easier than the analytical methods because we can simply take the average over the multiple realizations. For example, Ref.~\cite{TNJ2021} shows that the field-level modeling provides a fast way to compute the summary statistics and their covariance matrices incorporating survey window function due to non-trivial geometry and varying depth. In addition, Ref.~\cite{TNJ2022} presents the two-loop power spectrum and one-loop bispectrum of matter in redshift-space with this grid-based method. The possibility of field-level inference bypassing the summary statistics \cite{Andrews/etal:2022} further strengthens the motivation for the field-based method. 

In this {\it letter}, we present a novel $n$EPT ($n$th-order Eulerian Perturbation Theory) scheme for modeling the nonlinear density field. For the field-level SPT calculation, we use the GridSPT \cite{TNJ2018} that, unlike LPT, directly generates the density and velocity fields on grids without using particles. While using the recursion relations of the SPT to compute nonlinear fields at each order, $n$EPT differs from the other PT methods in computing the summary statistics: namely, $n$EPT first adds all nonlinear contributions to the density field up to the $n$th order, then compute the summary statistics. By contrast, the SPT computes the summary statistics by collecting the contributions at the fixed order in the linear density contrast $\delta_{\rm L}$.

In what follows, we show that $n$EPT models the nonlinear LSS with stunning accuracy, much better than the current state-of-the-art two-loop PT predictions.

{\it GridSPT and $n$EPT} --- 
For a given realization of the linear density field on regular grid points, the GridSPT \cite{TNJ2018} provides a way to compute the matter density field $\delta$ and the velocity field ${\bm v}$ of LSS perturbatively by solving the fluid equations:
\ba
&\dot{\delta} + \nabla\cdot\left[(1+\delta){\bm v}\right] = 0 ,
\\
&\dot{\bm v} + ({\bm v}\cdot\nabla){\bm v} + \frac{\dot a}{a}{\bm v} = - \nabla\phi ,
\ea
along with the Poisson equation:
\ba
\nabla^2 \phi = 4\pi G \bar{\rho}_\mathrm{m}\, a^2 \delta .
\ea
Here, dot represents the conformal-time derivative, $d\tau=dt/a$ with $a(t)$ being the scale factor and $t$ being the cosmic time, $\nabla$ is comoving-coordinate derivative, $\bar{\rho}_\mathrm{m}$ is the mean matter density, and $\phi$ is the peculiar gravitational potential. The set of equations describes the non-relativistic-matter (cold-dark matter and baryon) fluid on scales larger than the baryonic Jeans scale. 
Following the standard practice of SPT, we assume irrotational velocity and expand the density field and the normalized velocity-divergence field $\theta\equiv -\(\nabla\cdot{\bm v}\)/(aHf)$ as 
\ba
\delta(\tau,{\bm x})=&\sum_n [D(\tau)]^n\delta^{(n)}({\bm x}),
\label{eq:def_delta}
\\
\theta(\tau,{\bm x})=&\sum_n [D(\tau)]^n\theta^{(n)}({\bm x}).
\label{eq:def_theta}
\ea
Making use of the fast Fourier transform, the GridSPT enables us to quickly generate the $n$th order quantities $\delta^{(n)}$ and $\theta^{(n)}$ at each grid point following the configuration-space SPT recursion relation \cite{TNJ2018}. Here, $D$ denotes the linear growth factor and $f\equiv d\ln D/d\ln a$.

The crucial difference between $n$EPT and the usual PT is that in $n$EPT, we first compute the nonlinear density field in \refeq{def_delta} up to a fixed order $n$, then estimate the summary statistics, such as power spectrum and bispectrum, directly from $\delta$. For example, for $n=5$, the power spectrum from 5EPT reads
\ba
P_{\rm 5EPT}
=&
D^2 P_{11} 
+
2 D^3 P_{12}
+
D^4 \(2 P_{13} + 2 P_{22}\)
\vs
&+
D^5 \(2 P_{14} + 2P_{23}\)
+
D^6 \(2P_{15} + 2P_{24} + P_{33}\)
\vs
&+
D^7 \(2P_{25} + 2P_{34} \)
+
D^8 \(2P_{35} + P_{44}\) 
\vs
&+
2 D^9 P_{45} 
+ 
D^{10} P_{55} ,
\label{eq:nEPT}
\ea
which clearly differs from the nonlinear power spectrum in the usual PT:
\ba
P_{\rm PT}^{(2\text{-}{\rm loop})}
=&
D^2 P_{11} 
+ D^4 \(2P_{13} + P_{22}\)
\vs
&+ D^6 \(2P_{15} + 2P_{24} + P_{33}\) .
\label{eq:PT}
\ea
Here, we use the shorthand notation of
\be
\left\langle \delta^{(n)}({\bm k})\delta^{(m)}({\bm k}')\right\rangle\equiv(2\pi)^3 
P_{nm}(k)\delta^D({\bm k}+{\bm k'}) ,
\ee 
and suppress the $\tau$ and $k$ dependencies to avoid the clutter. The first (second) bracket in \refeq{PT} is called one-loop (two-loop) contribution in SPT. 

{\it N-body simulations} --- 
We test the performance of the $n$EPT modeling of the nonlinear power spectrum by comparing the $n$EPT prediction in \refeq{nEPT} against a series of $N$-body simulations. First, we use the baseline $N$-body simulation in Ref.~\cite{TNJ2018}: $1024^3$ particles in $L_\mathrm{box}=1\,\mathrm{Gpc}/h$ box with flat-$\Lambda$CDM cosmology ($\Omega_\mathrm{m}=0.279$, $h=0.701$, $n_s=0.96$, $\sigma_8=0.8159$) consistent with WMAP 5-year results \citep{komatsu2009five}. Then, we use the $N$-body simulation results from the Dark Quest project~\cite{DQ1} aiming to model the cosmological dependence of halo and matter statistics in the six-parameter $w$CDM cosmologies. The $20$ simulations are for the $20$ test cosmologies arranged uniformly over the six-dimensional hyperrectangle based on a maxi-min distance Latin hypercube design. In particular, we use the high-resolution suite with $2048^3$ mass elements in $(1\,\mathrm{Gpc}/h)^3$ periodic comoving boxes.
For both cases, we measure the matter power spectrum employing $1024^3$ grid points for FFT, with the aliasing artifact and the cloud-in-cells mass assignment kernel corrected in Fourier space \cite{Jing05,Sefusatti16}. The measurement error is much less than $1\%$ up to the Nyquist frequency of $k=3.2\,h/\mathrm{Mpc}$.

{\it $P(k)$ comparison: $n${\rm EPT} vs. N-{\rm body}} ---
\begin{figure*}
    \centering
    \includegraphics[width=2\columnwidth]{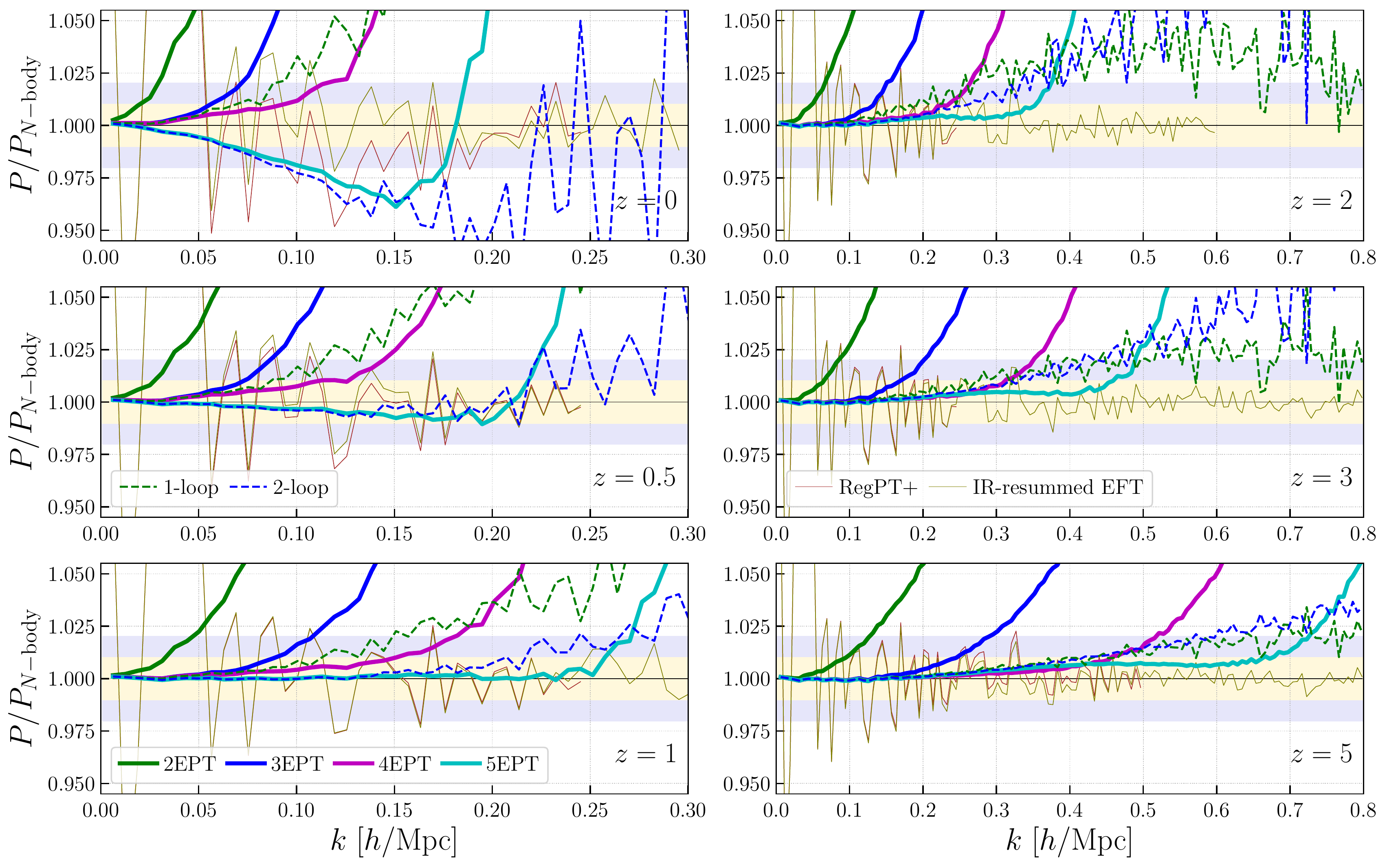}
    \caption{The ratios of the model power spectra to the $N$-body results for the baseline WMAP-5yr cosmology at redshifts $z = 0,\,0.5,\,1,\,2,\,3,$ and $5$. The thin dashed lines are the one-loop (green), and two-loop (blue) power spectra from SPT calculations. The thick solid lines are the result from $n$EPT calculations: 2EPT (green), 3EPT (blue), 4EPT (magenta) and 5EPT (cyan). Both SPT and $n$EPT results are measured from the density field in GridSPT using the same initial linear density field generating  the initial condition of the $N$-body simulation. The two thin solid lines are the \at{two-loop}  results of RegPT+ (brown) and IR-resummed EFT (olive) using the smooth (theory) linear power spectrum. The yellow and lavender bands indicate the $\pm$1\% and $\pm$2\% regions. We truncated RegPT+ and IR-resummed EFT beyond the $k_{\rm max}$ that gives rise to the minimum reduced $\chi^2$.}
    \label{fig:WMAPpk}
\end{figure*}
\begin{figure}[t]
    \centering
    \includegraphics[width=\columnwidth]{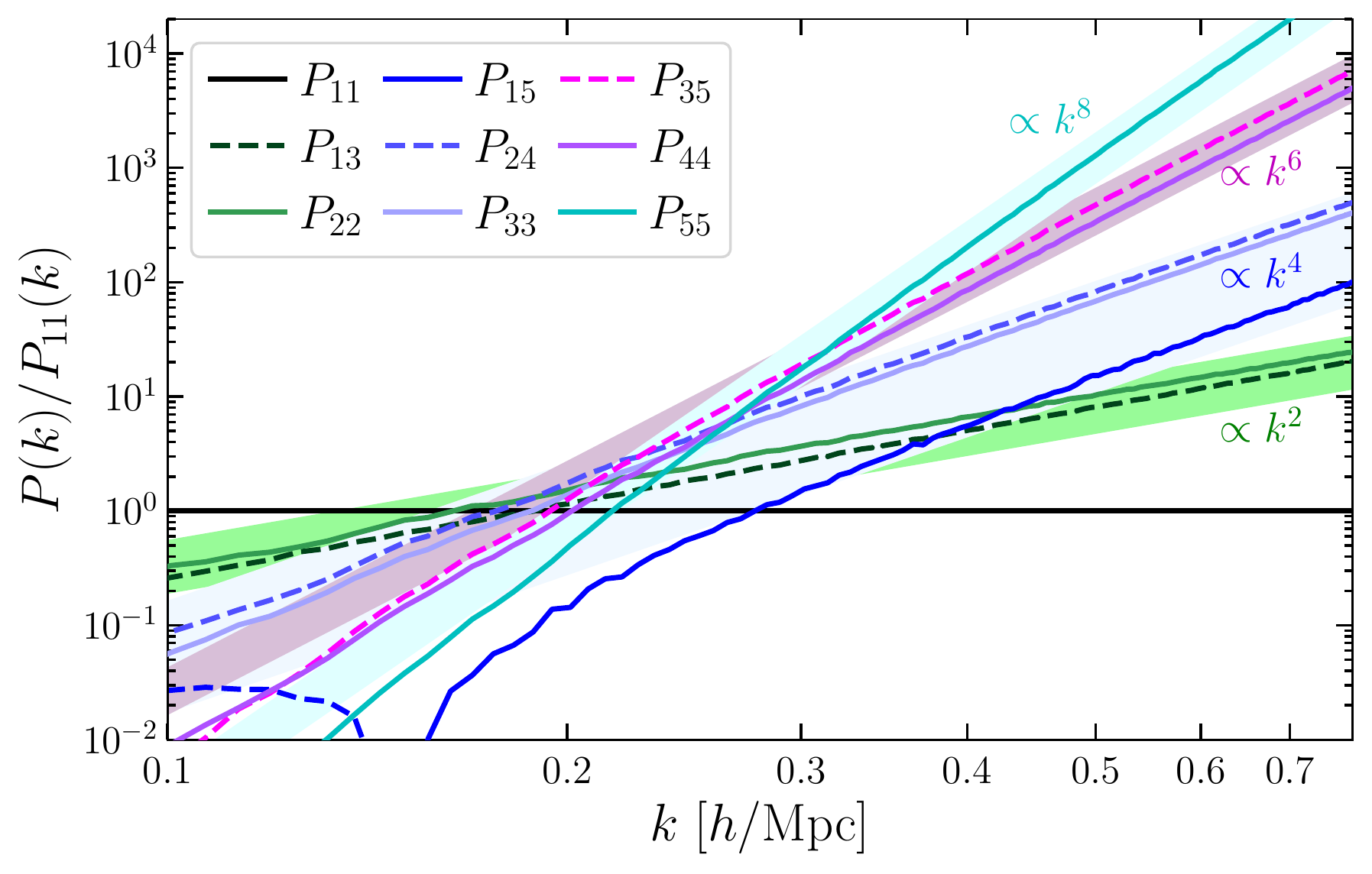}
    \caption{The ratios of even-order components $P_{nm}(k)$ to the linear power spectrum. 
    Solid (Dashed) lines denote positive (negative) values, and the components of the same order are plotted with similar colors. The shaded regions show the power law: $k^2$ (green), $k^4$ (blue), $k^6$ (magenta), and $k^8$ (cyan).}
    \label{fig:Pij}
\end{figure}
To make a face-to-face comparison with the $N$-body results, we calculate the GridSPT nonlinear density field using the same initial linear density field that generates the initial condition for corresponding $N$-body simulations.

When computing the Fourier-space quantities using real-space recursion relations, one must apply the cutoff to reduce the spurious impact from the small-scale (UV) modes. For the baseline calculation, we use the cut-off wavenumber $k_{\rm cut}^{\rm UV} = 256k_F = 1.61\, h/\mathrm{Mpc}$, but we shall also present the results with different $k_{\rm cut}$ later. Here, $k_F=2\pi/L_{\rm box}$ is the fundamental wavenumber. To avoid the aliasing effect, we adopt the generalized Orszag rule \cite{schmidt2021nLPT, TNJ2022} to zero-pad $k>2/(n+1)k_{\rm Nyquist}$ in linear density field for computing the $n$-th order field, where $k_{\rm Nyquist}=\pi N_{\rm grid}/L_{\rm box}$ is the Nyquist wavenumber with the one-dimensional grid size $N_{\rm grid}$. Requiring that $k_{\rm cut}^{\rm UV}<2/(n+1)k_{\rm Nyquist}$ sets the minimum $N_{\rm grid}$ that we use for the GridSPT calculation. For the baseline computation, we use $N_{\rm grid}={\rm 1536}$.

We have computed up to fifth-order GridSPT density fields that are sufficient for calculating SPT power spectrum to two-loop level and $5$EPT by using, respectively, \refeq{nEPT} and \refeq{PT}. \reffig{WMAPpk} shows the ratios of various model nonlinear power spectra to the baseline $N$-body power spectrum at following six redshifts: $z = 0,\,0.5,\,1,\,2,\,3,$ and $5$. Models plotted here are SPT (dashed lines), $n$EPT (thick solid lines), RegPT+ (thin brown line \cite{osato2019perturbation}), and IR-resummed EFT (thin olive line \cite{osato2019perturbation}). To facilitate the comparison, we highlight the one- and two-percentage ranges by yellow and lavender bands at the center and extend the three high-redshift (right) panels to $k=0.8~h/\mathrm{Mpc}$.

First, we note that the agreement between $n$EPT and $N$-body improves significantly as $n$ increases for $z \gtrsim 0.5$, and the 5EPT (the cyan lines) agrees with $N$-body results better than one percent to larger wavenumber than two-loop SPT $P(k,z)$ for $z \gtrsim 1$. Such accuracy of $5$EPT can only be matched with two-loop results of the RegPT+ and IR-resummed EFT that employ, respectively, one and three free parameters. Here, we show the RegPT+ and IR-resummed EFT to the maximum wavenumber minimizing the reduced $\chi^2$ assuming the diagonal covariance matrix with $\sigma[P(k)]=P(k)/\sqrt{N_k}$ \cite{jeong:thesis}. Note that for RegPT+ with $z\le3$, we find $k_{\rm max}=0.25~h$/Mpc. It is worth reminding the readers that $n$EPT requires no free parameters.

The $n$EPT and SPT results are also much smoother than the RegPT+ and IR-resummed EFT results. This is because the latter two models are the ensemble averages while $n$EPT and SPT are computed with the input linear density field of the $N$-body. In addition, we have added the odd-order terms (with odd power of $D$ in \refeq{nEPT}) to one-loop and two-loop power spectra. Although much smaller than the even-power terms, these odd-power terms indeed make the power spectrum from the same realization closer to the $N$-body result \citep{Takahashi_etal2008, TNJ2018}, especially for the large-scale Fourier modes \citep{TNJ2022}. 

We note the characteristic up-turn feature in the $n$EPT results that is absent in SPT \cite{jeong2006perturbation}; namely, $n$EPT does not \textit{suffer} from the poor convergence in SPT whose residual shows alternating-series-like behavior, which motivates the development of Renormalized PT \cite{crocce2006renormalized}. Instead, $n$EPT enjoys well-regulated high-$k$ behavior from the fact that contributions coming from higher order are stiffer by a factor of $k^2$, as we show in \reffig{Pij}.

Finally, one noteworthy feature in \reffig{WMAPpk} is that no wiggling feature appears in the ratio between $n$EPT and $N$-body around the Baryon Acoustic Oscillation (BAO) scales, which means that the damping of BAO has been accurately captured by $n$EPT, and the IR-resummation \citep{Senatore_Zaldarriaga2015, Vlah_etal2016} might not be necessary for $n$EPT. 

\begin{figure*}[t]
    \centering
    \includegraphics[width=2\columnwidth]{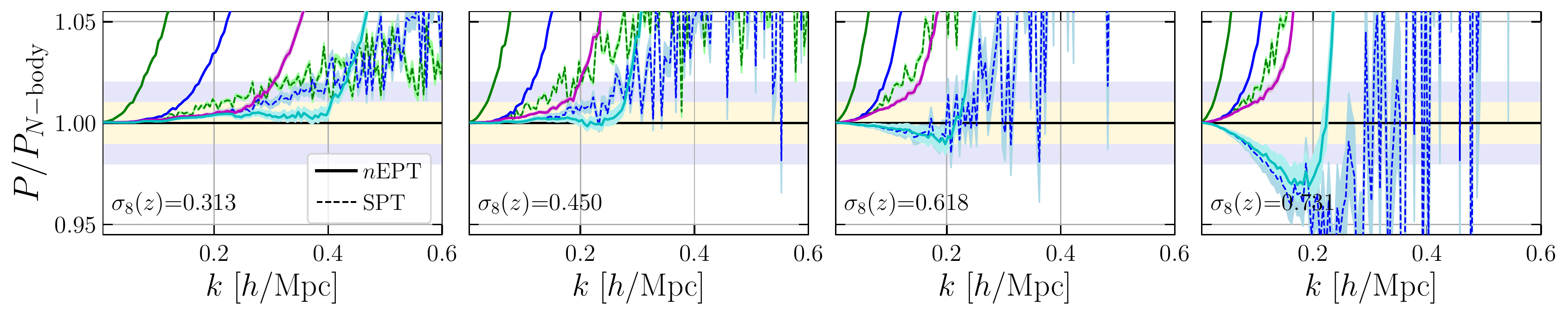}
    \includegraphics[width=2\columnwidth]{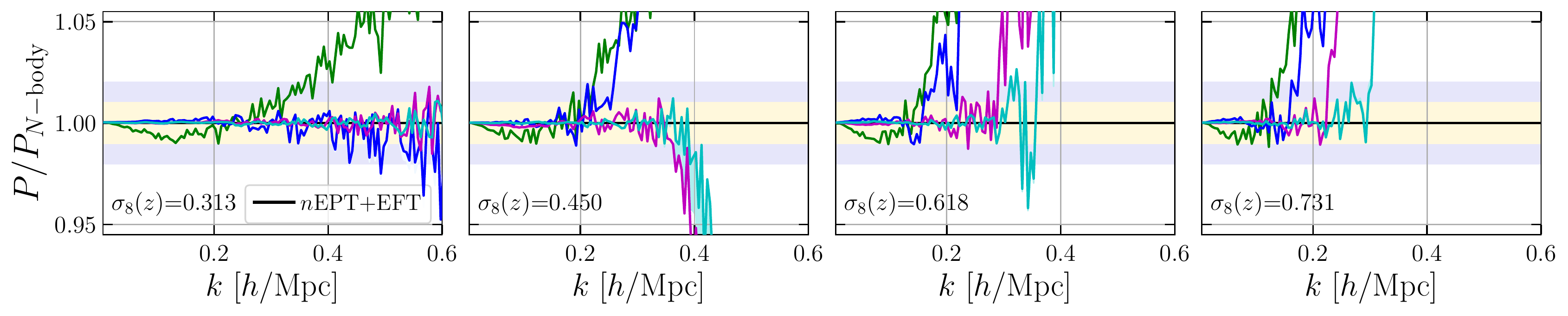}
    \caption{(\textit{Top}) The ratios of the real-space power spectra from $n$EPT (solid lines) and SPT (dashed lines) to the $N$-body results in four representative Dark Quest cosmologies. The colors are the same as that in \reffig{WMAPpk}. The shades show the range of $n$EPT power spectrum with different $k_{\rm cut}^{\rm UV}$ between $1.26 \,h/\mathrm{Mpc}$ and $2.14 \,h/\mathrm{Mpc}$. 
    (\textit{Lower}) The same as the Top panel but for the $n$EPT power spectrum with EFT correction in \refeq{EFTpk}.}
    \label{fig:darkquest}
\end{figure*}

\begin{figure*}
    \centering
    \includegraphics[width=\columnwidth]{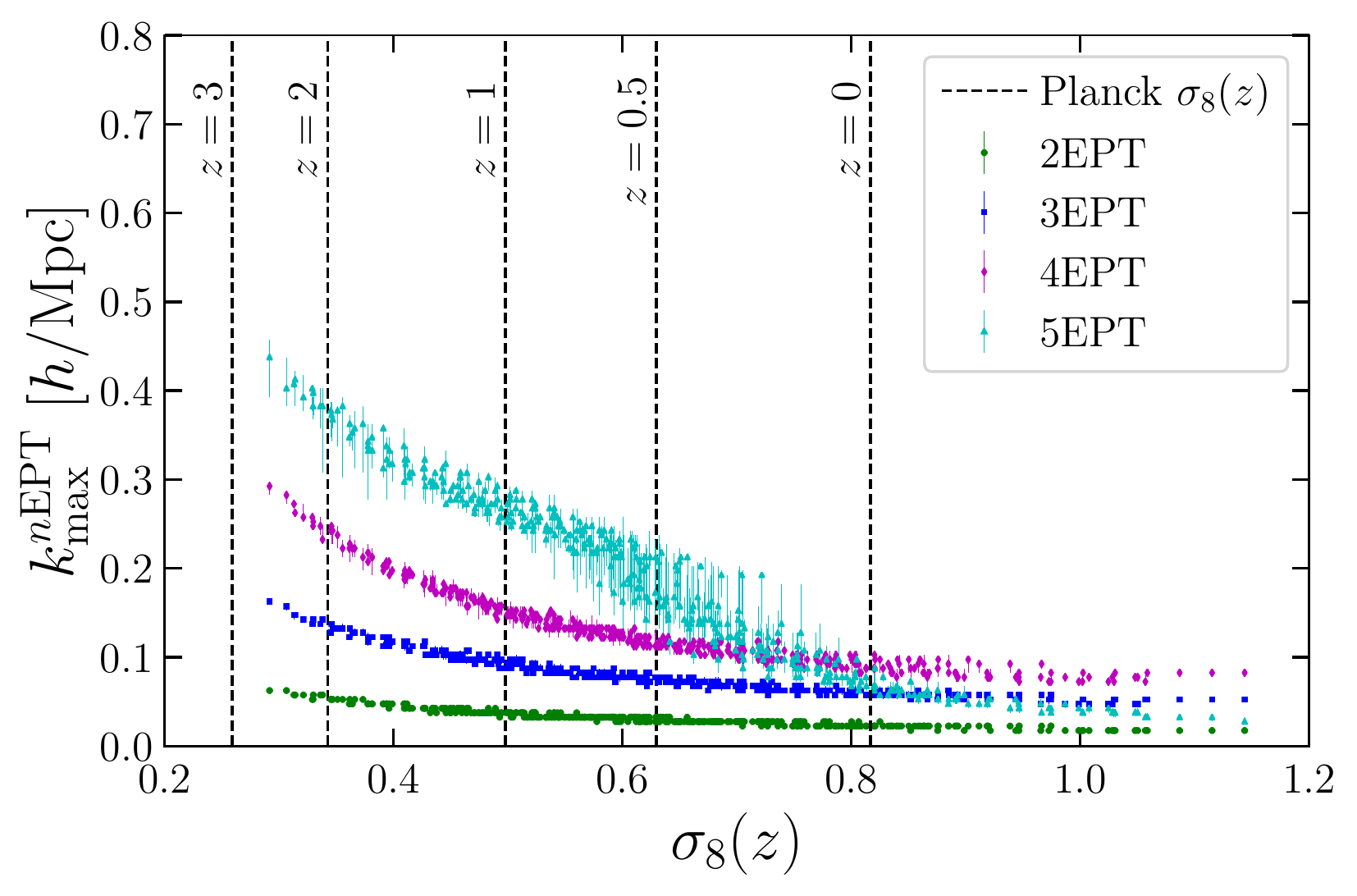}
    \includegraphics[width=\columnwidth]{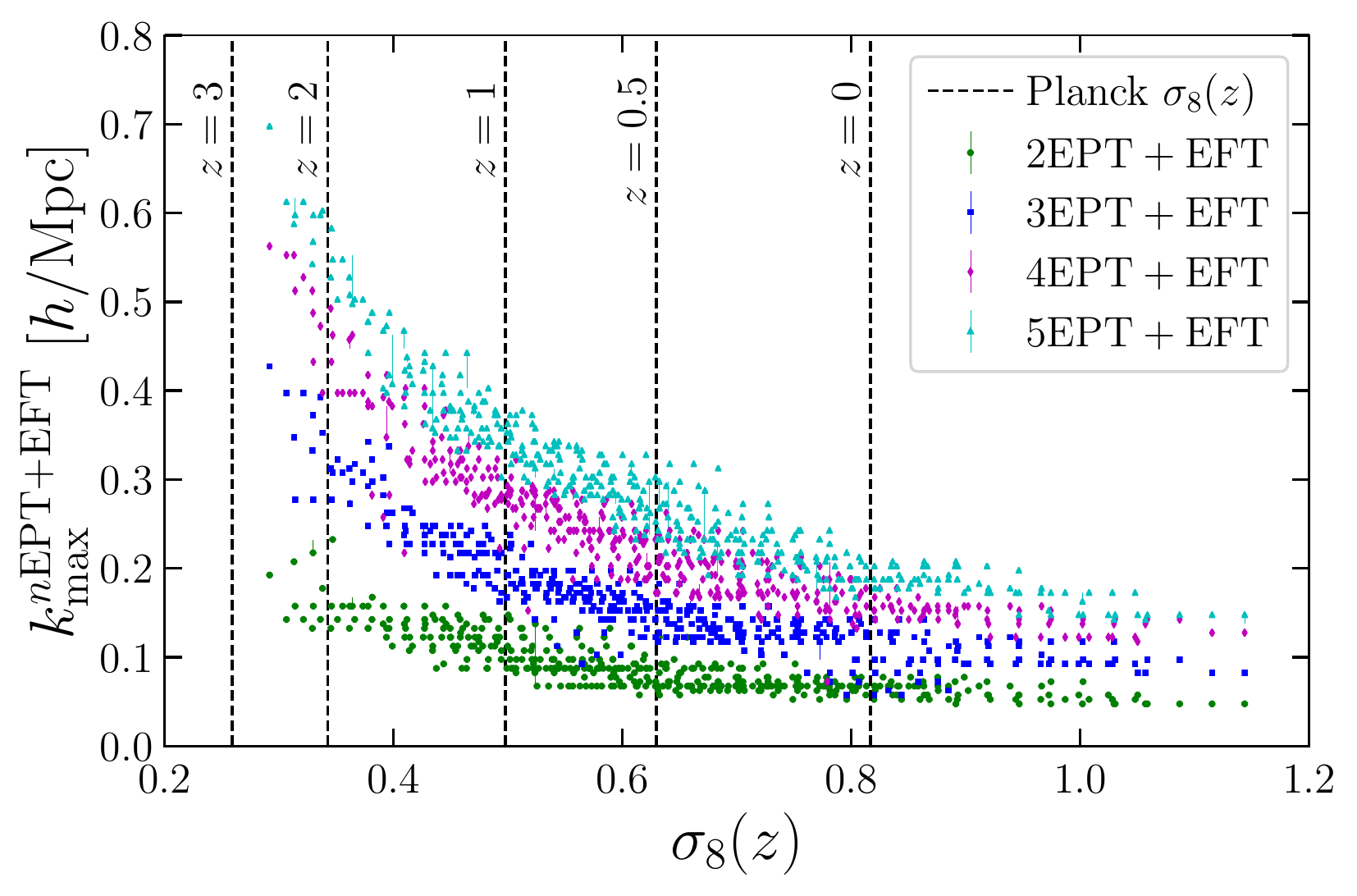}
    \caption{(\textit{Left}) The anti-correlation between the $k_{\max}$, the maximum wavenumber where $n$EPT matches N-body result to 1\% accuracy, of $n$EPT and $\sigma_8(z)$ for Dark-Quest simulation's all 20 cosmologies at 21 redshifts from $z = 0$ to $z = 1.48$. The error bars show the range of $k_{\max}$ with varying UV cutoff between $(1.26, 2.14)\, h/\mathrm{Mpc}$. The dashed lines indicate the value of $\sigma_8(z)$ in \textit{Planck} cosmology at redshifts $z = 0, 0.5, 1, 1.5, 2, 3$. (\textit{Right}) The same as the left panel but for the $n$EPT power spectrum with EFT correction in \refeq{EFTpk}.}
    \label{fig:kmax_s8}
\end{figure*}
We confirm that the same conclusion also holds for cosmological models different from the WMAP-5yr cosmology by comparing the $n$EPT results to the outcome from 20 Dark Quest simulations, at 21 redshifts from $z =0$ to $1.48$. Furthermore, we find that the $k_{\rm max}$, maximum wavenumber below which $n$EPT models the $N$-body result to one-percent accuracy, depends primarily on the $\sigma_8(z)=\sigma_8 D(z)$ value at the redshift. The top panel of \reffig{darkquest} shows four representative results with different $\sigma_8(z)$, and the left panel of \reffig{kmax_s8} shows the $k_{\rm max}$ measured from $n$EPT as a function of $\sigma_8(z)$. 

Here, we test the effect of the UV cutoff by calculating the $n$EPT power spectra with two other UV cutoffs, $(k_{\rm cut, 1}^{\rm UV}, k_{\rm cut, 2}^{\rm UV}) = (200, 340)k_F = (1.26, 2.14)\; h/$Mpc, and show the result as 
shaded regions in \reffig{darkquest}, and as ranges in \reffig{kmax_s8}. As expected, the higher-order $n$EPT is much more sensitive to the UV cutoff than the lower-order $n$EPT. Although going to 5EPT can significantly improve the accuracy of modeling the nonlinearities in matter clustering, for example, 5EPT is accurate up to $k_{\max} = 0.35\,(0.40)\;h$/Mpc at redshift $z = 2\,(3)$ in \textit{Planck} cosmology (dashed vertical lines in \reffig{kmax_s8}), one must be cautious on the UV sensitivity.

The UV-cutoff-dependence of $n$EPT, however, can be absorbed into the EFT-like counter terms. Motivated by \reffig{Pij}, we have included the EFT correction as
\bea
\label{eq:EFTpk}
\tilde P_{n{\rm EPT}}(k) = P_{n{\rm EPT}}(k) - \sum_{i=1}^{n-1} \alpha_i k^{2i} P_{\rm 11}(k),
\eea
where $P_{11}$ is the linear power spectrum of the $N$-body simulation, and $\{\alpha_i\}$ are free parameters that we fit from the measured power spectrum. As we have done for the RegPT+ and IR-resummed EFT, we find the $k_{\rm max}$ at which $\tilde P_{n{\rm EPT}}(k)$ proivdes the best fit to the $N$-body results. The shades in the lower panel of \reffig{darkquest} and the ranges in the right panel of \reffig{kmax_s8} are too narrow to be identified, which indicates that the EFT counterterms in \refeq{EFTpk} absorb the UV sensitivity in $n$EPT. Furthermore, as shown in the right panel of \reffig{kmax_s8}, the EFT correction improves the $k_{\max}$ of all $n$EPT power spectra significantly, especially at low $\sigma_8(z)$. For instance, with EFT correction, 5EPT can work accurately up to $k_{\max} = 0.6\,h/\mathrm{Mpc}$ at $z=2$.

{\it Conclusion} ---
In this {\it letter}, we present a novel $n$EPT resummation scheme and show that the $n$EPT outperforms one-loop and two-loop SPT as well as two-loop results of the RegPT+ and the IR-resummed EFT  without employing any free parameters. The resummation scheme also offers well-regulated convergence behavior at each successive $n$, bypassing the pathological behavior shown in SPT.

To be a successful theory for modeling observed galaxy clustering, the $n$EPT still needs to incorporate the galaxy bias and the redshift-space distortion, but we anticipate that $n$EPT must still thrive, at the very least, by following the prescriptions in SPT and EFTofLSS. However, taking advantage of having both density and velocity at each grid point, one can directly implement the non-linear redshift-space distortion mapping to improve the modeling accuracy further \cite{TNJ2022}. Upon the addition of galaxy bias and redshift-space distortion, the field-level modeling with $n$EPT will be a powerful data-analysis tool for future high-redshift galaxy surveys.	

Finally, while we have only demonstrated the accuracy of the $n$EPT scheme, a more in-depth theoretical study of the underlying reason for such behavior is desired.
\\
\begin{acknowledgments}
DJ is supported by KIAS Individual Grant PG088301 at Korea Institute for Advanced Study. This work was supported in part by MEXT/JSPS KAKENHI Grant Number JP19H00677 (TN), JP20H05861, JP21H01081 (AT and TN), JP21J00011, JP22K14036 (KO), and JP22K03634 (TN). We also acknowledge financial support from Japan Science and Technology Agency (JST) AIP Acceleration Research Grant Number JP20317829 (AT and TN). Numerical computations were carried out at the ROAR supercomputer at Penn State University, Yukawa Institute Computer Facility, and Cray XC50 at Center for Computational Astrophysics, National Astronomical Observatory of Japan.
\end{acknowledgments}

\appendix

\bibliography{PTbib}

\end{document}